\documentclass[aps,prl,preprintnumbers,twocolumn,nofootinbib,
,floatfix]{revtex4-1}
\usepackage[dvips]{graphicx}
\usepackage{bm,latexsym,amsmath,amssymb,amsfonts,mathrsfs}
\usepackage{color}

\input{colordvi.tex}

\newcommand{\im}{\mathrm{Im}}

\newcommand{\M}[0]{\mathcal{M}}

\newcommand{\Lag}[0]{\mathcal{L}}

\newcommand{\Landau}[0]{\mathcal{O}}

\newcommand{\dint}[0]{\mathrm{d}}

\newcommand{\mpl}[0]{M_{\text{Pl}}}

\newcommand{\phdm}{\sigma_{\gamma\text{-X}}}

\begin{document}

\preprint{KOBE-COSMO-22-5}

\title{
Phenomenological Motivation for Gravitational Positivity Bounds:
\\
A Case Study of Dark Sector Physics}

\author{Toshifumi Noumi}

\author{Sota Sato}

\author{Junsei Tokuda}

\affiliation{Department of Physics, Kobe University, Kobe 657-8501, Japan}

\begin{abstract}

Positivity bounds on scattering amplitudes provide a necessary condition for a low-energy effective field theory to have a consistent ultraviolet completion. Their extension to gravity theories has been studied in the past years aiming at application to the swampland program, showing that positivity bounds hold at least approximately even in the presence of gravity. A theoretical issue in this context is how much negativity is allowed for a given scattering process. We show that this issue is relevant to physics within the scope of ongoing experiments, especially in the context of dark sector physics. A detailed analysis of dark photon scenarios is provided as an illustrative example. 
Our results not merely show the phenomenological importance of the theoretical study of gravitational positivity bounds, but also open up an exciting possibility of exploring the nature of quantum gravity via experimental search of the dark sector. 
\end{abstract}

\maketitle
\section{I. Introduction}

The concept of effective field theory (EFT) and ultraviolet (UV) completion is a key framework for connecting various scales in nature. When applied to gravity, it will open up a road toward quantum gravity phenomenology. The swampland program~\cite{Vafa:2005ui} is aiming at this direction by exploring UV constraints on gravitational EFTs and their implications for particle physics and cosmology. See~\cite{Brennan:2017rbf, Palti:2019pca, vanBeest:2021lhn} for review articles.

Positivity bounds on scattering amplitudes is one of the interesting approaches to such UV constraints on EFTs. While the bounds were originally formulated in nongravitational theories with a mass gap~\cite{Pham:1985cr,Ananthanarayan:1994hf,Adams_2006}, their extension to gravity theories has been studied intensively, having application to the swampland program in mind~\cite{Hamada:2018dde,Bellazzini:2019xts,Alberte:2020jsk,Tokuda_2020,Loges:2020trf,Herrero-Valea:2020wxz, Caron-Huot:2021rmr,Alberte:2021dnj,Bellazzini:2021oaj,Chiang:2022jep}. A finding there is that positivity bounds hold at least approximately even in gravity theories.

For concreteness, we consider an $s$-$u$ symmetric scattering amplitude $\M(s,t)$ of the process $AB\to A B$ up to $\Landau(\mpl^{-2})$ and perform its infrared (IR) expansion as
\begin{eqnarray}
\M(s,t)&=&\sum_{n,m=0}^\infty\frac{c_{n,m}}{n!m!}\left(\frac{s-u}{2}\right)^nt^m
\nonumber
\\
&\quad&
+(s,t,u\text{-poles of graviton exchange})
\,,
\end{eqnarray}
where $(s,t,u)$ are the standard Mandelstam variables and they satisfy $s+t+u=2(m_A^2+m_B^2)$. Then, the positivity bound implies
\begin{equation}
\label{bound_c20}
c_{2,0}\geq \frac{\lambda}{M_{\rm Pl}^2M^2}
\quad(\lambda=\pm1)
\,,
\end{equation}
where the sign $\lambda$ and the scale $M$ depend on details of UV completion of gravity. See the next section for the definitions of $\lambda$ and $M$ in terms of gravitational Regge amplitudes at high energies and also the assumptions used to derive the bound~\eqref{bound_c20}.

When the threshold energy $m_{\rm th}$, i.e., the mass of the lightest intermediate on-shell state, is below the UV cutoff scale $\Lambda$ of the EFT, it is convenient to define~\cite{Bellazzini:2016xrt, deRham:2017imi}
\begin{equation}
B^{(2)}(\Lambda) 
    :=
    c_{2,0}
	 -
    \frac{4}{\pi} \int_{m_{\rm th}^2}^{\Lambda^2} \dint s \frac{\im\, \M (s,0)}{(s-m_{A}^2-m_B^2)^3}\,,\label{Bdef}
\end{equation}
which is calculable within the EFT. The bound~\eqref{bound_c20} on $c_{2,0}$ is now sharpened as
\begin{equation}
\label{bound_B2}
B^{(2)}(\Lambda) \geq \frac{\lambda}{M_{\rm Pl}^2M^2}\,.
\end{equation}

In the past years, the bounds~\eqref{bound_c20} and~\eqref{bound_B2} have been used in the swampland program and it turned out that the sign $\lambda$ and the scale $M$ are crucial for deriving the swampland conditions. For example, the sign $\lambda$ is crucial in the context of the weak gravity conjecture~\cite{Arkani-Hamed:2006emk} for black holes, which implies that the charge-to-mass ratios of extremal black holes are increased by higher derivative corrections. In particular, negativity of $c_{2,0}$ for a certain helicity amplitudes of light-by-light scattering implies violation of the conjecture. Hence, it is crucial to understand for which scattering process and in what class of UV completion of gravity, negativity of $c_{2,0}$ is allowed. See, e.g.,~\cite{Kats:2006xp,Cheung:2018cwt,Hamada:2018dde,Bellazzini:2019xts,Jones:2019nev,Loges:2019jzs,Cano:2019oma,Cano:2019ycn,Loges:2020trf,Cremonini:2020smy,Arkani-Hamed:2021ajd,Cano:2021nzo,Henriksson:2022oeu,Aalsma:2022knj} for recent discussion.

The scale $M$ is also important to use the bound \eqref{bound_B2} for constraining the particle spectrum in gravitational EFTs~\cite{Cheung:2014ega,Andriolo:2018lvp, Chen:2019qvr,Alberte:2020jsk, Alberte:2020bdz,Aoki:2021ckh,Noumi:2021uuv,Alberte:2021dnj}.
As we explain later, if the scale $M$ is at a UV scale $M\gtrsim \Lambda$, one may derive nontrivial constraints generically. On the other hand, if the sign $\lambda$ is negative and the scale $M$ is sufficiently low, the bounds are trivially satisfied without giving any constraint on IR physics. Interestingly, a recent paper \cite{Alberte:2021dnj} pointed out that for a certain helicity amplitude of graviton-photon scattering, the sign $\lambda$ has to be negative and also the scale $M$ has to be an IR scale for compatibility with the sum rule. Since the sign $\lambda$ and the scale $M$ depend on each scattering process and details of its UV completion, it is of great interests how generic this feature is beyond the graviton-photon scattering.\footnote{For example, if negativity of $c_{2,0}$ cannot be ruled out
in light-by-light scattering eventually, this provides a counterexample for
the black hole weak gravity conjecture. So far, there is no known
counterexample in string theory, so this issue would tell us if the
conjecture is specific in string theory or more generic in quantum
gravity.}

In this paper we address importance of this rather technical issue by demonstrating that it is relevant to physics within the scope of ongoing experiments, especially in the context of dark sector physics. This implies not only the phenomenological importance of further studies on gravitational positivity bounds, but also the exciting possibility of exploring the nature of quantum gravity via experimental search of the dark sector.  In this study, we analyze dark photon scenarios as an illustrative example.

\section{II. Gravitational Positivity Bounds}

We begin by a brief review of positivity bounds in the presence of gravity. A set of sufficient conditions for deriving the bounds~\eqref{bound_c20} and~\eqref{bound_B2} are the following:
\begin{enumerate}
    \item [(i)] {\it Unitarity}: The imaginary part of $\M$ is nonnegative, $\im\,\M(s\geq m_\text{th}^2,0)\geq0$. 
   
    \item [(ii)] {\it Analyticity}: $\M$ is analytic on the physical sheet of the complex $s$-plane except for poles and discontinuities on the real axis demanded by unitarity.
    \item [(iii)] {\it Mild behavior in the Regge limit}: The high-energy behavior of $\M(s,t)$ is sufficiently mild to satisfy $\lim_{|s|\to\infty} |\M (s,t<0)/s^2|=0$.
    \item [(iv)] {\it Regge behavior}: The $s^2$ bound (iii) is satisfied as a consequence of Reggeization by an infinite tower of higher-spin states.
\end{enumerate}
The properties (i)-(iii) are satisfied in the standard UV complete theory with a mass gap. Also, they are satisfied by construction in perturbative string theory (which emerged in the $S$-matrix theory context). Even though it is nontrivial if these properties should hold in general gravity theories,\footnote{See, e.g.,~\cite {Haring:2022cyf} for recent discussion on
high-energy behavior of gravitational scattering.}
we assume that (i)-(iii) hold for a small negative $t<0$ up to $\Landau(\mpl^{-2})$. By contrast, the property (iv) is naturally derived from the properties (i)-(iii) in the process with massless spin-2 exchange.
For our purpose, it is convenient to parameterize the imaginary part of the Regge amplitude up to $\Landau(\mpl^{-2})$ as 
\begin{equation}
	\im\,\M(s,t)
	\simeq 
	f(t) 
	\left(
		s/M_*^2
	\right)^{\alpha(t)}
	+
	\cdots
	\,\,
	(s\geq M_*^2)
	\,.\label{regge1a}
\end{equation} 
Here $M_*$ is the Reggeization scale, which is well above the mass scale of the higher-spin states, e.g., the string scale $M_s$ in perturbative string theory. Also, we suppressed terms irrelevant to Reggeization of graviton exchange.

Now the twice-subtracted dispersion relation for $t<0$ reads
\begin{eqnarray}
c_{2,0}+\Landau(t)
&=&
\frac{2}{M_{\rm Pl}^2t}
+\frac{4}{\pi}\int_{M_*^2}^{\infty} \dint s 
		\frac{f(t)\left(
		s/M_*^2
	\right)^{\alpha(t)}}{(s+t/2-m_A^2-m_B^2)^3}
		\nonumber
			\\
		&\quad&
+\frac{4}{\pi}
\int_{m_{\rm th}^2}^{M_*^2} \dint s 
		\frac{\im\, \M (s,t)}{(s+t/2-m_A^2-m_B^2)^3}
\,,
\end{eqnarray}
where the singular graviton $1/t$-pole must be canceled with the second term on the right-hand side (RHS). 
The cancellation fixes the value of $f(0)$.  Also, evaluating the $\Landau(t^0)$ term gives~\cite{Tokuda_2020}
\begin{equation}
B^{(2)}(\Lambda)=\frac{4}{\pi}
\int_{\Lambda^2}^{M_*^2} \dint s 
		\frac{\im\, \M (s,0)}{(s-m_A^2-m_B^2)^3}
		+\frac{\lambda}{M_{\rm Pl}^2M^2}
  \label{eq:sumrule1}
\end{equation}
with $\lambda\,(=\pm1)$ and $M\,(>0)$ defined by
\begin{equation}
\frac{\lambda}{M^2}=-
\left[\frac{2f'(0)}{f(0)}-\frac{\alpha''(0)}{\alpha'(0)}\right]\,.\label{Mdef}
\end{equation}
Then, $\im\, \M (s,0)\geq0$ implies the bounds~\eqref{bound_c20} and~\eqref{bound_B2}.

\section{III. General consideration}\label{sec:general}

We discuss general implications of the bound \eqref{bound_B2} for the dark sector physics. For earlier discussions along this line of considerations, see, e.g., \cite{Andriolo:2018lvp,Alberte:2021dnj}.
Let us consider the following model in which the Standard Model Lagrangian $\Lag_\text{SM}$ and the Lagrangian of the dark sector $\Lag_\text{DS}$ are included in addition to the Einstein-Hilbert term: 
\begin{equation}
	\Lag
	=
	\frac{\mpl^2}{2}R
	+ \Lag_\text{SM}[A_b,\cdots]
	+ \Lag_\text{DS}[X,\cdots]
	\,. \label{genlag}
\end{equation}
Here, $A_b$ and $X$ represent an ordinary photon and a hidden particle in the dark sector, respectively. We define $\Lag_\text{DS}$ such that direct interactions between the dark sector and the Standard Model, if exist, are included in $\Lag_\text{DS}$. In this section, we assume that the matter sector described by $\Lag_\text{SM}+\Lag_\text{DS}$ is renormalizable and neglect nonrenormalizable terms for simplicity. 
We will add nonrenormalizable terms to the matter Lagrangian and discuss the implications in the next section. We  discuss the implication of \eqref{bound_B2} for the scattering amplitude $\M(s,t)$ of the process $\gamma X\to \gamma X$. We will later consider the case $X=\gamma'$, the dark photon as an illustrative example.

We can calculate $B^{(2)}(\Lambda)$ by evaluating the amplitude $\M$ within EFT in eq.~\eqref{Bdef}.
We decompose diagrams up to $\Landau(\mpl^{-2})$ into two parts: the nongravitational diagrams in which gravitons are absent and those with a graviton exchange. We refer to the contributions from the former diagrams and the latter diagrams as $B^{(2)}_{\text{non-grav}}(\Lambda)$ and $B^{(2)}_{\text{grav}}(\Lambda)$, respectively.
Because the matter sector itself is renormalizable, the nongravitational EFT amplitude $ \M_\text{non-grav}$ satisfies the twice-subtracted dispersion relation, leading to  
\begin{equation}
	B^{(2)}_\text{non-grav}(\Lambda)
	=
	\frac{4}{\pi}\int^\infty_{\Lambda^2}\mathrm{d}s\,
	\frac{\im\,\M_\text{non-grav}(s,0)}{(s-m_X^2)^3}
	\,,\label{formula1}
\end{equation} 
where $m_X$ is the mass of the particle $X$. $\M_\text{non-grav}(s,0)$ at $s>\Lambda^2$ is the nongravitational EFT amplitude extrapolated to high energy regimes $s>\Lambda^2$ by analytic continuation.

By contrast, the GR amplitude does not satisfy the twice-subtracted dispersion relation because of the nonrenormalizable nature of GR.
An electron one-loop correction to the diagram with graviton $t$-channel exchange 
gives rise to the $\Lambda$-independent $B^{(2)}_\text{grav}(\Lambda)\supset -\Landau (e^2\mpl^{-2}m_e^{-2})$ while the corresponding $s,u$-channel diagrams give $\Landau (e^2\mpl^{-2}\Lambda^{-2})$ terms. 
 Then, $B^{(2)}_\text{grav}(\Lambda)$ is given by a negative constant in a good approximation at $\Lambda\gg m_e$:~\footnote{Strictly speaking, loop corrections to the $XXh$-vertex
could also contribute to $B^{(2)}_\protect \text {grav}$. The values of such
contributions depend on the details of $\protect \mathcal {L}_\protect \text
{DS}$, the dark sector model. However, as long as $X$ is a stable particle,
the sign of such contributions would be negative as far as we know. Hence,
adding such terms will not change the essence of the discussion.}
\begin{equation}
	B^{(2)}_\text{grav}(\Lambda)
	=
	- \Landau\left(\frac{e^2}{\mpl^{2}m_e^2}\right)
	\,.\label{bgravity}
\end{equation}
As a result, the bound \eqref{bound_B2} reads 
\begin{equation}
    B^{(2)}_{\text{non-grav}}(\Lambda)
	 \geq
	 \Landau\left(e^2\mpl^{-2}m_e^{-2}\right) - \lambda\mpl^{-2}M^{-2}
    \label{pos4}
    \,.
\end{equation}
The RHS depends on the unknown scale $M$ given in \eqref{Mdef}: a priori even the sign of the RHS is undetermined. {\it Implications of \eqref{pos4} crucially depend on the value of $M$.} If $M$ is given by an IR scale as $M\lesssim m_e/e$, the RHS of \eqref{pos4} may become negative. In this case, the bound \eqref{pos4} is satisfied in any renormalizable unitary matter theories coupled to gravity because of the condition $B^{(2)}_\text{non-grav}(\Lambda)>0$, which follows from the formula \eqref{formula1}.  

If $M$ is given by the scale of UV physics such as the mass of higher-spin states and satisfies the condition $M\gg m_e/e$, we can then ignore the second term on the RHS of \eqref{pos4}, leading to
\begin{equation}
    \qquad\,\,\,
	 B^{(2)}_{\text{non-grav}}(\Lambda)
    \geq - B^{(2)}_{\text{grav}}(\Lambda)
	 =
	 \Landau\left(\frac{e^2}{\mpl^{2}m_e^2}\right)
    \label{pos5}
    \,.
\end{equation} 
This gives a nontrivial bound on the EFT even though the matter sector is renormalizable. In particular, this implies an upper bound on $\Lambda$ in terms of parameters in the model such as coupling constants, because $B^{(2)}_\text{non-grav}(\Lambda\to\infty)\to0$ as explicitly seen from Eq.~\eqref{formula1}. 

Using the formula \eqref{formula1} and the relation $\im\,\M_\text{non-grav}(s,0)\simeq s\phdm^\text{non-grav}(s)$ at $s\gg m^2_X$, \eqref{pos5} becomes\footnote{Based on analogous discussions, we can derive nontrivial bounds on processes with
external Higgs bosons. See~\cite
{Alberte:2020jsk,Noumi:2021uuv} for discussions of gravitational positivity
bounds on scalar scatterings.}
\begin{equation}
	\frac{4}{\pi}\int^\infty_{\Lambda^2}\mathrm{d}s\,\frac{\phdm^\text{non-grav}(s)}{s^2}
	\geq
	-B^{(2)}_\text{grav}
	=
	\Landau\left(\frac{e^2}{\mpl^{2}m_e^2}\right)
	\,.\label{genimp1}
\end{equation}
Here $\phdm^\text{non-grav}(s)$ is a total cross section of the $\gamma\text{-}X$ scattering evaluated from the non-gravitational EFT amplitude $\M_\text{non-grav}$ at the center-of-mass energy $\sqrt{s}$. 
Therefore, the bound \eqref{pos5} implies that a particle $X$ in the dark sector must nongravitationally couple to photon to satisfy the lower bound \eqref{genimp1} on $\phdm^\text{non-grav}$, meaning that {\it the dark sector cannot be too dark}.\footnote{In the standard renormalizable theories, the cross section is consistent with the Froissart-Martin bound. The integral in x \ref{genimp1} may be then dominated by its lower end.
Assuming this, (\ref{genimp1}) implies a
lower bound on $\sigma^\text{non-grav}_{\gamma \protect \text {-X}}(\Lambda ^2)$ as 
\begin{equation} 
  \sigma^\text{non-grav}_{\gamma \text {-X}}(\Lambda ^2) \gtrsim
\mathcal{O}\left(\frac {e^2\Lambda ^2}{M_{\text{Pl}}^{2}m_e^2}\right) .
\label{general1} 
\end{equation} 
This could also be useful for understanding the
implication of the bound (\ref{pos5}) for given
models in terms of the total cross section.}
This is reminiscent of typical dark sector scenarios in extra dimensions, where decoupling of the two sectors is realized by taking a large volume limit and so it cannot be achieved as long as the volume is finite to make the four dimensional Planck mass $M_{\rm Pl}$ finite. 
In the next section, we study implications of \eqref{pos5}-\eqref{genimp1} more concretely for dark photon models.


For completeness, let us point out that even if $M$ is given by the IR scale, there remains a possibility that we get nontrivial bounds similar to \eqref{pos5}-\eqref{genimp1}: for instance, it is possible that the RHS of \eqref{pos4} is still given by some positive quantity of $\mathcal{O}(e^2\mpl^{-2}m_e^{-2})$ and consequently \eqref{pos5}-\eqref{genimp1} are derived even when we have $M\sim m_e/e$. It is also possible that the first term on the RHS of \eqref{eq:sumrule1} which must be positive thanks to the unitarity, plays a role to give the bounds \eqref{pos5}-\eqref{genimp1}.

It is important to study the scale $M$ and \eqref{pos5}-\eqref{genimp1} in explicit quantum gravity scenarios. For instance, \cite{Hamada:2023cyt} pointed out an interesting string compacification example in which a resultant 4d effective theory has $\mathcal N=2$ supersymmetry and states charged under $U(1)$ gauge symmetry. The charged states can be arbitrarily light so that the bounds \eqref{pos5}-\eqref{genimp1} are not obviously satisfied. It would be interesting to study such a nontrivial setup in light of gravitational positivity bounds.

\section{IV. Dark photons}
The dark photon is the gauge boson $A_a^\mu$ of an extra $U(1)$ symmetry in the dark sector to which we refer as $U(1)_\text{d}$~\cite{Holdom:1985ag} (see~\cite{Fabbrichesi:2020wbt} for a review). The kinetic mixing of $A_a$ and the ordinary photon $A_b$ is renormalizable, so that there is no reason to exclude it. Indeed, it appears naturally if there exist heavy fields which are charged under both of the electromagnetic $U(1)$ and the hidden $U(1)_\text{d}$. Also, $A_a$ is massive in general. We then consider the massive dark photon model described by \eqref{genlag} with 
\begin{equation}
	\Lag_\text{DS}
	=
	- \frac{1}{4} (F_{a,\mu\nu})^2 
	- \frac{1}{2}m_{A'}^2 (A_{a,\mu})^2
	- \frac{\epsilon}{2} F_{a,\mu\nu}F_{b}^{\mu\nu}
	+ \Lag_\text{hm}
	\, ,\label{dslag}
\end{equation}
where $\Lag_\text{hm}$ denotes Lagrangian of matters in the dark sector.\footnote{To be precise, the dark photon kinetically mixes with the hyper-charge gauge boson. In this case, the dark photon couples to the weak neutral current as well. However, this coupling is suppressed as $\mathcal{O}\left(m_{A'}^{2}/m_{Z}^{2}\right)$ and irrelevant in the parameter space studied in the present paper.}
The kinetic terms are diagonalized in terms of new fields $(A,A')$ which are defined by
\begin{eqnarray}
    \begin{pmatrix}
        A_a \\
        A_b \\
    \end{pmatrix}
    &=
    \begin{pmatrix}
        1/\sqrt{1-\epsilon^2} & 0\\
        -\epsilon/\sqrt{1-\epsilon^2} & 1\\
    \end{pmatrix}
    \begin{pmatrix}
        A' \\
        A \\
    \end{pmatrix}
    \label{rot1}
    \,.
\end{eqnarray}
The rotation \eqref{rot1} generates tiny couplings between $A'$ and the Standard Model: they are included in $\Lag_\text{SM}|_{A_b\to A-\left(\epsilon/\sqrt{1-\epsilon^2}\right)A'}$. By contrast, couplings between $A$ and hidden matters are not induced by this rotation. Below, we call $A$ and $A'$ photon and the dark photon, respectively, and analyze the photon-dark photon scattering $\gamma\gamma'\to\gamma\gamma'$. 

There are two possibilities for the photon-dark photon scattering: one where the dominant diagrams involve Standard Model particles, and another where the dominant diagrams involve particles from the dark sector.
We consider implications of gravitational positivity bounds on each of these two possibilities respectively.

The amplitude depends on the helicity configurations in general. For the process with transversely polarized dark photon, we consider 
\begin{eqnarray}
      \M_T (s,t) := &\frac{1}{4}\bigl[ \M (1^+ 2^+ 3^+ 4^+)+\M (1^+ 2^- 3^+ 4^-)
                 \nonumber\\
					 &+\M (1^- 2^- 3^- 4^-)+\M (1^- 2^+ 3^- 4^+)\bigr]
                 \,, \nonumber
\end{eqnarray}
where 1 and 2 (3 and 4) are the ingoing (outgoing) photon and dark photon, respectively, and the superscript $\pm$ denotes the helicity. For the process with longitudinal modes, we consider 
\begin{equation}
     \M_L (s,t) :=  \frac{1}{2}\left[\M (1^+ 2^L 3^+ 4^L)+\M (1^- 2^L 3^- 4^L)\right]
                 \,,\nonumber
\end{equation}  
where $L$ means the longitudinal polarization. For these amplitudes, the crossing symmetry implies the $s\leftrightarrow u$ permutation invariance.

Below, we refer to $(B^{(2)}_\text{non-grav},\,B^{(2)}_\text{grav})$ for $\M_T$ and $\M_L$ as $(B^{(2)}_\text{T,non-grav},\,B^{(2)}_\text{T,grav})$ and $(B^{(2)}_\text{L,non-grav},\,B^{(2)}_\text{L,grav})$, respectively.

The one-loop diagrams are calculated by using the Mathematica packages FeynRules~\cite{Christensen:2009jx,Christensen:2008py,Alloul:2013bka}, FeynArts~\cite{Hahn:2000kx},
FeynCalc~\cite{Mertig:1990an,Shtabovenko:2020gxv,Shtabovenko:2016sxi} and Package-X~\cite{Patel:2015tea}.

\subsection{A simplest model}
To begin with, let us consider the simplest setup where the contributions of heavy particles including those in the dark sector to $B^{(2)}\left( \Lambda \right)$ are negligibly small.
In this setup, the scattering process $\gamma\gamma'\to\gamma\gamma'$ is described by the Lagrangian \eqref{dslag}~ with $\Lag_\text{hm}=0$ and it is mediated only by the Standard Model particles. Since new particles coupled to photon below 1 TeV have not been found, we choose $\Lambda\geq1$ TeV.

\begin{figure}[tbp]
    \centering
    \includegraphics[scale=0.22]{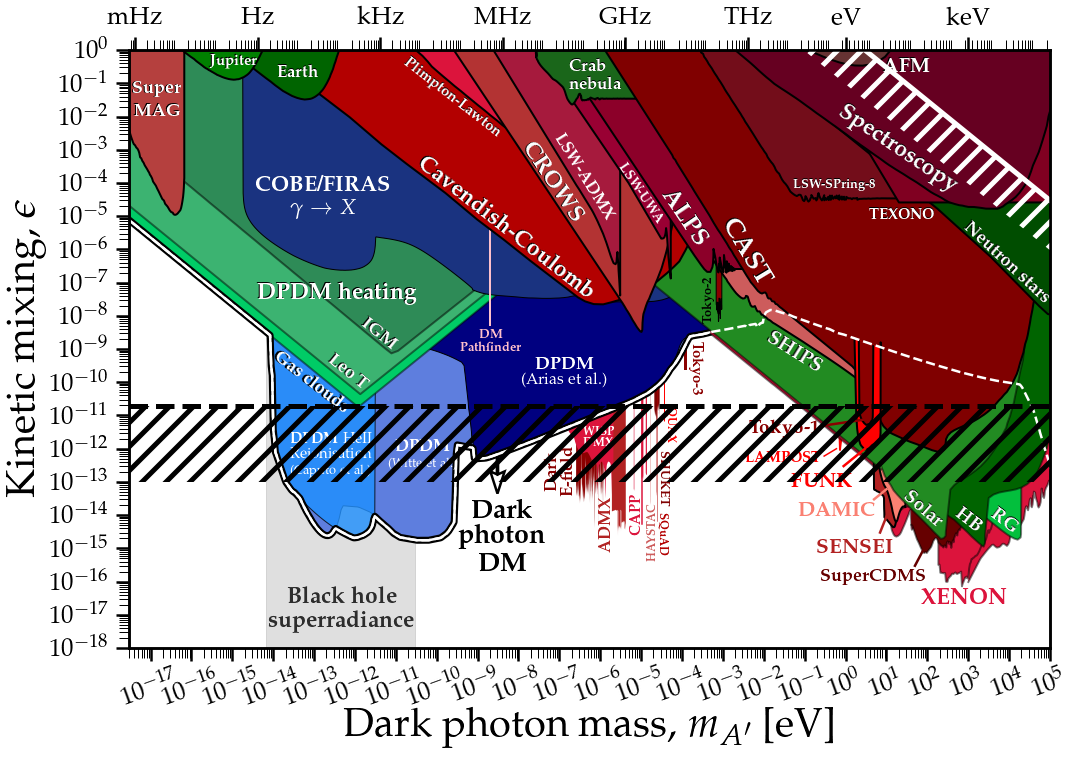}
    \caption{The upper bounds \eqref{Tbound1} and \eqref{Lbound1} for $\Lambda=1$ TeV are plotted in the $(m_{A'},\epsilon)$-plane by the black dotted line and the white line, respectively. 
    The figure is made based on the data given in~\cite{AxionLimits,Caputo:2021eaa}. 
    We find that the bounds are so strong and can be tested against experiments. This also implies the possibility of experimentally exploring general properties of quantum gravity S-matrix by testing the bounds. 
    }
	\label{fig:limitt1}
\end{figure}
The form of interactions between $A'$ and the Standard Model particles are the same as those for $A$ except that the coupling strength is suppressed by a factor $\epsilon/\sqrt{1-\epsilon^2}\simeq\epsilon$. 
As a result, the calculations of $\M_T$ and $\M_L$ are similar to those of the amplitude of the light-by-light scattering which has been done in~\cite{Aoki:2021ckh}. As it is the case for the light-by-light scattering, in our case, the dominant contributions to $B^{(2)}_{\text{non-grav}}(\Lambda)$ at $\Lambda\geq 1\text{ TeV}$ are given by the W-boson loop whereas the dominant term in $B^{(2)}_{\text{grav}}$ arises from the electron, the lightest charged particle in the Standard Model. The results are,
\begin{eqnarray}
	&B^{(2)}_\text{T,non-grav}(\Lambda)&
	\simeq 
	\frac{32 \alpha ^2 \epsilon^2}{m_W^2 \Lambda ^2}
	\,,
	B^{(2)}_\text{L,non-grav}(\Lambda)
	\simeq 
	\frac{8 \alpha^2 \epsilon^2 m_{A'}^2}
    { \Lambda ^2 m_W^4}
	\,,\nonumber\\
	&B^{(2)}_\text{T,grav}(\Lambda)&
	\simeq B^{(2)}_\text{L,grav}(\Lambda)
	\simeq
	 -\frac{11\alpha}{180\pi m_e^2 \mpl^2}
	\,,\label{nongrav_simplest}
\end{eqnarray}
where $\alpha\simeq1/137$ denotes the fine-structure constant. 
Then, \eqref{pos5} implies a lower bound on $\epsilon$: from $\M_T$ we obtain 
\begin{equation}
    \epsilon
	 \geq
	 \sqrt{\frac{11}{5760 \pi  \alpha }}\frac{m_W \Lambda}{m_e \mpl}
    \simeq
    1.9 \times 10^{-11}\times
    \left(\frac{\Lambda}{1\,\mathrm{TeV}}\right)
    \,,\label{Tbound1}
\end{equation}
while we obtain a stronger constraint from $\M_L$ as
\begin{eqnarray}
    \epsilon
	 &\geq&
	 \sqrt{\frac{11}{1440 \pi  \alpha }}\frac{m_W^2 \Lambda}{m_{A'} m_e \mpl}
\nonumber
\\
    &\simeq&
    3.0\times 10^{-3}\times
    \left(\frac{\Lambda}{1\,\mathrm{TeV}}\right)
    \left(\frac{1\,\mathrm{keV}}{m_{A'}}\right)
    \,.\label{Lbound1}
\end{eqnarray}
These bounds \eqref{Tbound1} and \eqref{Lbound1} are plotted in FIG.~\ref{fig:limitt1}.  
The bound \eqref{Lbound1} also shows the existence of {\it a lower bound on
$m_{A'}$} for given $\epsilon$ and $\Lambda$. 
Note that \eqref{Lbound1} only applies to the massive dark photon case. The origin of the lower bound on $m_{A'}$ is the suppression factor $(m_{A'}/m_W)^2$ contained in $B^{(2)}_\text{L,non-grav}$. This factor appears because longitudinal modes of dark photons are decoupled from W-bosons in the limit $m_{A'}\to0$.

We have neglected QCD contributions so far. Since the QCD scale is lower than the electroweak scale,  it may give relevant contributions to $B^{(2)}_\text{non-grav}$. 
Indeed, QCD effects, particularly the Pomeron exchange, provide the leading contribution to $B^{(2)}_\text{non-grav}$ in the Standard Model analysis of \cite{Aoki:2021ckh}.
If we extend the analysis to transversely polarized dark photons simply by multiplying the factor $\epsilon$ to the coupling between photon and Pomeron, the bound \eqref{Tbound1} is slightly relaxed as $\epsilon\gtrsim10^{-13}$ for $\Lambda\sim1$ TeV.

It would be interesting to investigate how much the bound \eqref{Lbound1} is relaxed when QCD is taken into account. In general, the QCD contributions to $\M_L$ will be suppressed by the dark photon mass $m_{A'}$ at least as long the mass is the St\"{u}ckelberg mass. We thus expect that the existence of a lower bound on $m_{A'}$ like \eqref{Lbound1} is a robust result, leaving further studies for future work.

\subsection{Remarks on higher-dimensional operators}

In the above analysis, we have ignored contributions from heavy particles. In general, however, we can also consider dark photon models in which new particles other than $\gamma'$ are also present whose mass scales are well below the quantum gravity scales such as $\mpl$ or $M_s$. Let us discuss how our bounds \eqref{Tbound1} and \eqref{Lbound1} on the kinetic mixing can be changed once new particles are introduced. 

Let us introduce a new particle $Y$ with mass $m_Y$ for illustrations. 
Below the mass scale $m_Y$, we can take into account the contribution from $Y$ to the process $\gamma\gamma'\to\gamma\gamma'$ by adding the higher-dimensional operators to the Lagrangian: operators which are relevant for the discussion of positivity bounds at the leading order are, 
\begin{align}
    \frac{C_1}{m_Y^4} F^2 F'^2 \ , 
    \quad
    \frac{C_2}{m_Y^6} F_{\mu\nu}F^{\nu}_{\ \rho}\ \partial^{\mu}F'_{\alpha\beta} \partial^{\rho}F'^{\alpha\beta}
    \,. \label{higher}
\end{align}
Here, Wilson coefficients $C_1$ and $C_2$ are typically given by the couplings between $Y$ and $(\gamma,\gamma')$. The contributions of the leading higher-dimensional operators \eqref{higher} to $B^{(2)}_\text{T,non-grav}(\Lambda)$ and $B^{(2)}_\text{L,non-grav}(\Lambda)$ are estimated as
\begin{subequations}
\label{higher_estimate}
\begin{align}
&    \left.
    B^{(2)}_\text{T,non-grav}(\Lambda)
    \right|_\text{operators \eqref{higher}}
    \sim \frac{C_1}{m_{Y}^4}\,,
    \\
&    \left.
    B^{(2)}_\text{L,non-grav}(\Lambda)
    \right|_\text{operators \eqref{higher}}
    \sim \frac{C_2 m_{A'}^2}{m_Y^6}\,,
\end{align}
\end{subequations}
where we have $\Lambda<m_Y$ because the cutoff scale cannot be heavier than the mass of $Y$ which is integrated out. When these contributions are larger than those from particles in the Standard Model \eqref{nongrav_simplest}, the implications of the positivity bound \eqref{pos5} will be different from  \eqref{Tbound1} and \eqref{Lbound1}.\footnote{The contribution from $Y$ to $B^{(2)}_\text{T,grav}$ and $B^{(2)}_\text{L,grav}$ will be typically $\mathcal{O}(\mpl^{-2}m_Y^{-2})$ which is negligible compared to those from electron \eqref{nongrav_simplest}.} According to eqs.~\eqref{nongrav_simplest} and \eqref{higher_estimate}, the contributions from higher-dimensional operators to $B^{(2)}_\text{T,non-grav}(\Lambda)$ and $B^{(2)}_\text{T,non-grav}(\Lambda)$ are  negligible when the following condition is satisfied, respectively:
\begin{subequations}
\label{criteria}
\begin{align}
    &m_Y
    \gtrsim
    \left(\frac{|C_1|}{\epsilon^{2}\alpha^2}\right)^{\frac{1}{4}}
    \sqrt{m_W\Lambda}
    \,,\\
    &m_Y
    \gtrsim
     \left(\frac{|C_2|}{\epsilon^{2}\alpha^2}\right)^{\frac{1}{6}}
    \left(m_W^2\Lambda\right)^{\frac{1}{3}}
    \,,
\end{align}
\end{subequations}
where we impose $(m_W<)\Lambda<m_Y$.

When Y carries only electromagnetic $U(1)$ charge, $C_1$ and $C_2$ are suppressed by the kinetic mixing:
\begin{align}
	C_1,\ C_2 \lesssim \mathcal{O}(\epsilon^2) \ .
\end{align}
As a result, the conditions \eqref{criteria} are always satisfied. Hence, the bounds \eqref{Tbound1} and \eqref{Lbound1} are robust against the addition of heavy particles charged only under $U(1)$.

On the other hand, $C_1$ and $C_2$ are not suppressed by $\epsilon$ in general if $Y$ is charged under both of $U(1)$ and $U(1)_\text{d}$. In particular, we can consider the scenario in which $|C_1|,|C_2|\gg \epsilon^2$ such that the conditions \eqref{criteria} are violated even for very heavy $m_Y$. 
Therefore, it will be possible to construct models which are consistent with the positivity bound \eqref{pos5} regardless of the value of $\epsilon$, by introducing heavy bi-charged particles.

\subsection{Adding bi-charged particles}

Now we construct a model in which both of the bound \eqref{pos5} and the current observational bound on $\epsilon$ are satisfied simultaneously. For simplicity, we set $\epsilon=0$ and instead add to the model two bi-charged vector bosons $(V^\mu_1,V^\mu_2)$ with the same charges $(q_\text{em},q_\text{em})$ under $U(1)$ but the opposite charges $(q_\text{d},-q_\text{d})$  under $U(1)_\text{d}$. 
We choose $q_\text{em}=q_\text{d}=1$ and write the $U(1)_\text{d}$ gauge coupling  as $e'$ below. We assume that their masses $m_{V_1}$ and $m_{V_2}$ are approximately identical: $m_V:=m_{V_1}\simeq m_{V_2}$. With this choice, these particles generate a tiny kinetic mixing suppressed by a factor $|m_{V_1}^2-m_{V_2}^2|/m_V^2\ll1$ so that the model can be consistent with the current observations. Since bi-charged particles are coupled to photon with coupling strength $e$, they must be sufficiently heavy to be consistent with experiments: we then impose $m_V\gtrsim 1 \text{ TeV}\gg m_e$. 

In this model, $B^{(2)}_\text{T,nongrav}(\Lambda)$ and $B^{(2)}_\text{L,nongrav}(\Lambda)$ at $\Lambda\gg m_V$ are evaluated up to one-loop level as   
\begin{eqnarray}
    B^{(2)}_{\text{T,non-grav}}(\Lambda)
	\simeq
	\frac{4e^2e'^2}{\pi^2m_V^2\Lambda^2}
	\,,\\
	B^{(2)}_{\text{L,non-grav}}(\Lambda)
	\simeq
	\frac{e^2e'^2m_{A'}^2}{\pi^2m_V^4\Lambda^2}
	\,.
\end{eqnarray}
Bi-charged particles also contribute to $B^{(2)}_\text{grav}$, but it is negligible compared to the electron contribution as long as $m_V\gg m_e$ and  $e'\lesssim 1$.   
The bound \eqref{pos5} is\footnote{Note that, even if we add bi-charged particles with spin
smaller than one, the bound will not be relaxed comparing with 
\eqref{Tbound2} and \eqref{Lbound2} because its contributions to
$B^{(2)}_\protect \text {T,non-grav}$ and $B^{(2)}_\protect \text
{L,non-grav}$ are proportional to $\Lambda ^{-4}$ and $m_{A'}^2\Lambda
^{-6}$, respectively.}
\begin{eqnarray}
	\Lambda
	&\leq&
	24 \sqrt{\frac{5}{11}}\frac{m_e}{m_V}e'\mpl
	\nonumber
	\\
	&\simeq&
	2.0\times 10^{13}
	\,\,\text{GeV}
\times	e'
		\left(
			\frac{1\,\text{TeV}}{m_V}
		\right)
	\label{Tbound2}
\end{eqnarray}
for $\M_T$, while for $\M_L$ it reads
\begin{eqnarray}
	\Lambda
	&\leq&
	12 \sqrt{\frac{5}{11}}\frac{m_em_{A'}}{m_V^2}e'\mpl
	\nonumber\\
	&\simeq&
	10
	\,\,\text{TeV}
	\times e'
		\left(
			\frac{m_{A'}}{1\,\text{keV}}
		\right)
		\left(
			\frac{1\,\text{TeV}}{m_V}
		\right)^2
	\,.\label{Lbound2}
\end{eqnarray}
Eq.~\eqref{Tbound2} shows that, by adding sufficiently light bi-charged spin-one particles appropriately, we can make massless dark photon models with a tiny kinetic mixing $\epsilon$ consistent with \eqref{pos5} even for large cutoff $\Lambda\gg1\,\text{TeV}$. When the dark photon is massive $m_{A'}>0$, the scattering of longitudinal modes gives a stronger constraint \eqref{Lbound2}. The constraint implies $m_{A'}\gtrsim e'$ keV because of the condition $\Lambda, m_V\gtrsim 1$ TeV. As discussed below \eqref{Lbound1}, this is due to the suppression of couplings between longitudinal modes and matters at small $m_{A'}$. 

We conclude that {\it dark photon cannot be too light}, at least in models we discussed. 
It would be interesting to explore models that accommodate a very light dark photon without violating the bound~\eqref{pos5}, leaving it for future work.

\section{V. Conclusion}

In this paper we addressed importance of determining the sign $\lambda$ and the scale $M$ in the gravitational positivity bound~\eqref{bound_B2}. If $\lambda$ is negative and $M$ is as small as the electron mass scale, the bound is trivially satisfied. By contrast, when $M$ is sufficiently large, the bound gives a nontrivial constraint on IR physics even if the sign $\lambda$ is negative. This gives a general bound~\eqref{genimp1} on the total cross section of the photon-dark particle scattering, which implies that the dark sector cannot be too dark.

We also provided more detailed analysis of dark photon scenarios. A general finding is that the scattering of longitudinal modes of dark photons gives a stronger constraint than that of transverse modes. In particular, dark photons cannot be too light. It would be interesting to explore its possible connection to the swampland argument on photon masses~\cite{Reece:2018zvv}. Besides, we observed that kinetic mixing with massive dark photons favors UV completion by bi-charged vector bosons whose masses are bounded by \eqref{Lbound2}, which is reminiscent of gauge unification.

We stress that the bounds obtained under the assumptions (i)-(iv) and $M\gg m_e/e$ are within the scope of ongoing dark photon searches.\footnote{Notice that even if the energy scale $M$ is an IR scale $M\sim m_e/e$, nontrivial bounds may remain depending on the $\mathcal{O}(1)$ coefficient of the Regge parameters: see also discussions at the last paragraph of sec.~III.} This implies not only the phenomenological importance of further studies on gravitational positivity bounds in four dimensional spacetime but also the exciting possibility of testing general properties of quantum gravity S-matrix via experimental search of dark sector physics.

\medskip
\begin{acknowledgments}
\paragraph{Acknowledgments.}

We would like to thank Yoshihiko Abe, Katsuki Aoki, Kimihiro Nomura, Ryo Saito, Satoshi Shirai and Masahito Yamazaki for useful discussion.
T.N. is supported in part by JSPS KAKENHI Grant No.~20H01902 and No.~22H01220, and MEXT KAKENHI Grant No.~21H00075, No.~21H05184 and No.~21H05462.
S.S. is supported in part by JST SPRING Grant No.~JPMJFS2126.
J.T. is supported in part by JSPS KAKENHI Grant No.~20J00912 and No.~21K13922.
\end{acknowledgments}

\end{document}